\documentclass[useAMS,usenatbib]{mn2e}

\usepackage{graphicx}

\def\bi {\begin{itemize}}
\def\ei {\end{itemize}}
\def\intgr{\textit {INTEGRAL}}

\def\asca{\textit {ASCA}}
\def\xmm{\textit {XMM-Newton}}
\def\psrb{PSR~B1259-63} 
\def\ROSAT{\textit {ROSAT}}
\def\sax{\textit{Beppo}SAX}

\def\rxp {2RXP~J130159.6-635806}
\def\deg {$^\circ$}

\usepackage{color}
\definecolor{red}{rgb}{0.7,0,0}
\definecolor{blue}{rgb}{0,0,0.7}

\begin{document}

\title{Discovery and study of the accreting pulsar \rxp}


\author[M. Chernyakova et al.]{M. Chernyakova$^{1,2}$\thanks{E-mail:Masha.Chernyakova@obs.unige.ch}\thanks{M.Chernyakova 
is on leave from Astro Space Center of the P.N.~Lebedev Physical Institute,  Moscow, Russia}, A. Lutovinov$^{3}$,
 J. Rodriguez$^{4,1}$, M. Revnivtsev $^{5,3}$
 \\
$^{1}$INTEGRAL Science Data Centre, Chemin d'\'Ecogia 16, 1290 Versoix, Switzerland\\
$^{2}$Geneva Observatory, 51 ch. des Maillettes, CH-1290 Sauverny, Switzerland \\
$^{3}$ Space Research Institute,  84/32 Profsoyuznaya Street, Moscow 117997, Russia\\
$^{4}$ CEA Saclay, DSM/DAPNIA/Service d'Astrophysique (CNRS UMR 7158 AIM), 91191 Gif sur Yvette, France\\
$^{5}$ Max-Planck-Institute f\"ur Astrophysik,
              Karl-Schwarzschild-Str. 1, D-85740 Garching bei M\"unchen}

\date{Received $<$date$>$  ; in original form  $<$date$>$ }
\pagerange{\pageref{firstpage}--\pageref{lastpage}} \pubyear{2002}

\maketitle
\label{firstpage}

\begin{abstract}
We report on analysis of the poorly studied source \rxp\ at
different epochs with \asca, \sax, \xmm\ and \intgr. 
The source shows coherent X-ray pulsations at a period $\sim 700$~s
with an average spin up rate of about $\dot{\nu}\sim 
2\times 10^{-13}$ Hz s$^{-1}$.
A broad band (1-60 keV) 
spectral analysis of \rxp\ based on almost simultaneous \xmm\ and
\intgr\ data demonstrates that the source has a spectrum typical of an 
accretion powered X-ray pulsar, i.e. an absorbed power law with a high
energy cut-off with a  photon index $\Gamma\sim 
0.5-1.0$ and a cut-off energy of $\sim 25$ keV. The long term behaviour of the
source, its spectral and timing properties, tend to indicate a  high mass X-ray binary with Be companion.
We also report on the 
identification of the likely infrared counterpart to \rxp. The interstellar 
reddening does not allow us to strongly constrain the spectral type of the 
counterpart.  The latter is, however, consistent with a Be star, 
the kind of which is often observed in accretion powered X-ray pulsars.
\end{abstract}

\begin{keywords}
pulsars: individual: \rxp~ --
gamma rays: observations -- X-rays: binaries -- X-rays: individual: \rxp~
 \end{keywords}

  
%



\maketitle


\section{Introduction}

On  February 7, 2004, during a routine Galactic plane scan, the
\textit{INTEGRAL} observatory detected a source which was not in the
\intgr\ reference catalog. Search in the archive led to the
identification of several \ROSAT\ sources in the \intgr\ error box.
Among them \rxp~ is the closest one  to the best estimate of
the source position obtained with \intgr\ \citep{chern-at}. The only
mention of this source in the literature before the observations
reported here (besides the \ROSAT\ catalog) can be found in
\citet{kaspi95}. In this paper the authors report results of \asca\
observations of the famous binary system \psrb, and mention the
presence of the another source located only 10 arc-minutes away to the
north-west.  \citet{kaspi95} note, in particular, that the absorption
column $N_{\mathrm{H}}$ in the direction of this source is higher than
that of \psrb, and that during their observations the brightness of \rxp~
was smaller, but comparable to that of \psrb.

We started to follow the source after its detection with
\textit{INTEGRAL} and our first \textit{XMM-Newton}
observation in a set of observations organized to monitor \psrb~
during its 2004 periastron passage (\psrb\ has a very long, 3.4 years, orbital
period).  Analyzing these observations we have noticed that \rxp,
which was also in the field of view, was significantly brighter than
it was during the \textit{ASCA} observations \citep{chern-at} \footnote{In the
telegram by a misprint 1RXP catalog was mentioned instead of 2RXP one}.  On January 
24, 2004 the 1-10 keV intensity was found to be approximately an order 
of magnitude higher than during the ASCA observation performed on August 13, 1995.

In 2001 -- 2004 \psrb\ was regularly monitored by \xmm\, and \rxp\ was
always in the \xmm\ field of view.  In this paper we present the analysis of all available
X-ray data from \asca, \sax, \intgr, and \xmm\ in order to understand the
nature of this variable source and investigate its properties.  In
particular, using the \xmm\ data we refine and improve the X-ray position, and
discover X-ray pulsations. We study the long term spectral evolution,
and through a simultaneous fit to the \xmm\ and \intgr\ data, show that
the hard X-ray source seen in \intgr\ and \rxp\ are very likely to be
the same object.\\ 
\indent The paper is organized as follows: in
Section 2 we present the sequences of observations and methods used
for data reduction and analysis.  In section 3 we present the results
obtained, and discuss them in Section 4. We then give a summary of our
analysis in the last part of the paper.

\section[]{Observations and Data Analysis}

\subsection{\intgr\ observations}

Since the launch of \intgr\ \citep{winkler} on  October 17, 2002, \rxp\
was several times in the field of view of the main instruments  
during the routine Galactic plane scans and pointed observations 
(see Table \ref{intdata} for details).  Most of the times the distance of the source 
from the center of the field of view  was too big, and
the exposure too short to use either the X-ray monitor JEM-X, or the 
spectrometer SPI. Therefore IBIS/ISGRI \citep{lebrun} is the only
instrument we can use in our analysis of this source. In this analysis
we have used the version 4.2 of the Offline Science Analysis (OSA)
software distributed by ISDC \citep{courvoisier}.

In 2003 the source was only marginally detected with IBIS/ISGRI, while in
the beginning of 2004 it was clearly seen in the $20-60$ keV energy
range. To obtain better results for spectral analysis we have
combined data obtained on January 26 and February 7, when source was the 
brightest.

\begin{table}
\caption{Journal of the \intgr\  observations of \rxp. \label{intdata}}
\begin{tabular}{c|c|c|c|c}
\hline
 Data & Date & MJD& Exposure   &20-60 keV Flux \\
  Set  &     & &      (ks)      &$10^{-11}$ergs s$^{-1}$cm$^{-2}$\\
\hline
I1& 2003-05-29 -- & 52788 --   &260&    1.4   $\pm$ 0.4\\
  & 2003-07-18 & 52838    &	 &      \\
I2& 2004-01-17&53021&5.3&3.83  $\pm$ 2.39\\
I3& 2004-01-26&53030&6.1&12.86 $\pm$ 2.83\\
I4& 2004-02-07&53042&4.2&15.04 $\pm$ 2.83\\
I5& 2004-02-19&53054&4.0&6.42  $\pm$ 3.10\\

\hline
\end{tabular}
\end{table}

\subsection{\xmm\ observations \label{xmmobs}}

In the course of  monitoring of \psrb, \xmm\ observed \rxp~ 10 times during
2001 -- 2004, see Table \ref{data} for the journal of the  observations. 
These data are a combination of public and private
observations. The Observation Data Files (ODFs) were obtained from the
online Science
Archive\footnote{http://xmm.vilspa.esa.es/external/xmm\_data\_acc/xsa/index.shtml},
and were then processed and filtered using {\sc
xmmselect} within the Science Analysis Software ({\sc sas}) v6.0.1.
In all the observations the source was observed with MOS1 and MOS2 detectors
only.  The 2001 -- 2003 observations (X1 -- X5) were done in the Full
Frame Mode, while the 2004 observations  were performed in the Small Window
Mode, in order to minimize pile-up problems for the primary source \psrb. 
In all observations a medium filter was used.
\begin{table}
\caption{Journal of  \asca\, \sax\, and  \xmm\ observations of \rxp. \label{data}}
\begin{center}
\begin{tabular}{c|c|c|c}
\hline
Data& Date & MJD& Exposure \\
 Set&      &    &  (ks)         \\
\hline
\multicolumn{4}{c}{\asca~}\\
\hline

A1&  1993-12-28   &  49349.28 & 20.7      \\
A2&  1994-01-26   &  49378.79 & 18.2      \\
A3&  1994-02-28   &  49411.61 &  8.3      \\
A4&  1995-02-07   &  49756.08 & 17.5      \\
A5&  1995-08-13   &  49942.98 & 19.6       \\
 \hline
\multicolumn{4}{c}{\sax}\\
\hline
S1&     1997-09-08 &50699.44       &    84        \\
\hline
\multicolumn{4}{c}{\xmm~}\\
\hline
X1&  2001-01-12 & 51921.73 & 11.3 \\
X2&  2001-07-11 & 52101.31 & 11.6 \\
X3&  2002-07-11 & 52467.24 & 41.0 \\
X4&  2003-01-29 & 52668.27 & 11.0 \\
X5&  2003-07-17 & 52837.53 & 11.0 \\
X6&  2004-01-24 & 53028.79 &  9.7 \\
X7&  2004-02-10 & 53045.43 &  5.2  \\
X8&  2004-02-16 & 53051.39 &  7.7 \\
X9&  2004-02-18 & 53052.02 &  5.2 \\
X10& 2004-02-20 & 53055.82 &  6.9 \\                                     
\hline
\end{tabular}
\end{center}
\end{table}

The spectra and light curves were extracted from a 
35$^\prime$$^\prime$ radius circle around the source position for the weak
state of the source (i.e. obs. X1 -- X5, X9, X10), and from a 
50$^\prime$$^\prime$ radius circle for the outburst phase (obs. X6 -- X8).  As
\rxp\ was not a primary goal of the \xmm\ observations its position is
shifted to the edge of the field of view,  and the shape of the source is
slightly elongated.  Therefore, in order to avoid mixing of source and
background photons for the weak states of the source, we collected background 
light curves and spectra  from a 35$^\prime$$^\prime$ radius circle located close 
to the source. For the bright state of the source we have used a circle
of larger radius, and  collected background light curves and spectra from
a 100$^\prime$$^\prime$ outer radius annulus centered on the source position.
 
Obs. X2, X4, X6 and X9 were partially affected by soft proton
flares.  Since proton flares originate from the interaction of the
soft protons in the Earth's magnetosphere with the telescope, their
timing behaviour is supposed to have no periodic structure. Therefore,
 no filtering of the data was applied to the timing analysis, as 
was done for another new X-ray pulsar IGR/AX J16320-4752
\citep{lut05}. We have nevertheless eliminated obs. X4 and X9 from our study 
as in these data sets the influence of the soft proton flares was
especially strong. Arrival times of the photons have been corrected to
the Solar System barycenter.  The pulse period was searched with the
epoch folding technique \citep{leahy83}: we produced periodograms and
derived the best-fit period for each data set. Ten bins per any trial period were used. 
For the determination of the uncertainty of the source period we used the bootstrap method. We 
simulated a number of source "fictional" lightcurves, generating randomly (in accordance with 
the poissonian statistics of the counts) its flux in each lightcurve bin.
These lightcurves provided us the range of "best-fit" periods of the source pulsations, 
therefore giving us information about the period uncertainty. Errors given in the paper
represent a $1\sigma$ confidence level.



For the spectral analysis the periods of soft protons need to be
filtered out. To exclude them we extracted light curves above 10 keV 
with a hundred second binning and excluded all time bins in which the count
was higher than 1.5 cnt/s.

Data from MOS1 and MOS2 detectors were combined in both timing and
spectral analysis in order to achieve better statistics.

\subsection{\asca\ observations}

\rxp\ was in the \asca\ field of view during the dates listed in Table
\ref{data}. In our subsequent analysis we have used the data of both Gas Imaging 
Spectrometers (GIS 2 and 3). The data were analyzed with the help of the 
standard tasks of LHEASOFT/FTOOLS 5.2 package in accordance with 
the recommendations of the \asca\ Guest Observer Facility.

\subsection{\sax\ observations}

During 1997  \rxp\ was several times within the  field of view of
the instruments of the \sax\ observatory. Unfortunately flux of the source detected
by the MECS telescopes was strongly contaminated by instrumental 
features (e.g. ``strongback'', see \citealt{mecs}), and therefore detailed
analysis of the source spectrum is not possible. However, the data obtained
can still be used for timing analysis. For the data reduction we
used standard tasks of LHEASOFT/FTOOLS 5.2 package. We only present here the 
results of an observation performed on September 8, 1997, when the statistics was 
good enough to perform a pulse search.

\begin{figure}
\begin{center}
\includegraphics[width=8cm,angle=0]{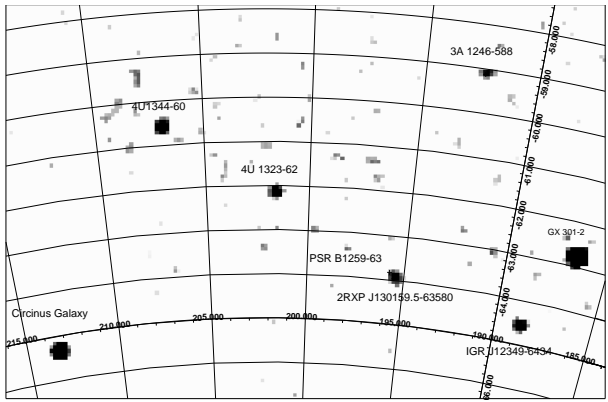}
\end{center}
\caption{20--60 keV significance mosaic of I1,I2,I3,I4, and I5
observations. Axis are in Equatorial J2000 coordinates (degrees).}
\label{intgr}
\end{figure}

\begin{figure}
\begin{center}
\includegraphics[width=8cm,angle=0]{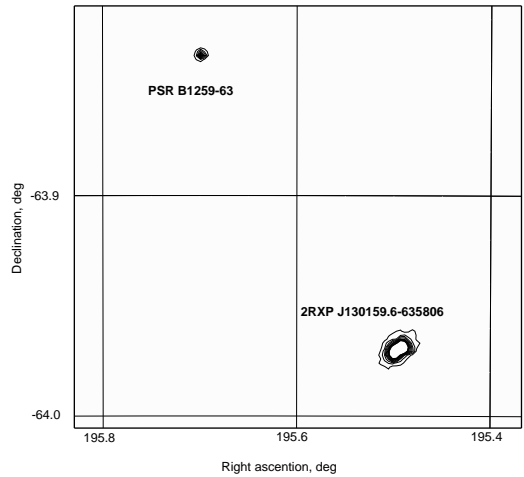}
\end{center}
\caption{Contour plot of \xmm\ field of view for the X6 observation. A
total of 10 contours were used with a linear scale. The external
contour corresponds to 5 counts per pixel, and the most internal one
to 50 counts per pixel. In this observation \rxp\ was forty times
brighter than \psrb.}
\label{xmm_ima}
\end{figure}

\section[]{Results}

\subsection{Imaging Analysis}
In Figure \ref{intgr} a zoom  of the mosaic of all the  \intgr\ observations
mentioned in Table \ref{intdata} is given. \rxp\ is clearly seen in the image,
along with a new source IGR J12349-6434 we have found
during our analysis (Chernyakova et al. 2005a). All sources shown in the image 
were taken into account  for a proper analysis of \intgr\ data (Goldwurm  A.
et al., 2003).

During the \xmm\ monitoring programme of \psrb,  two
sources were clearly detected (e.g. Fig. \ref{xmm_ima} represents the contour 
plot of Obs. X6).
Besides \psrb\ itself a second source can clearly be seen. The best coordinates we 
derive are RA$_ {J2000}$=13$^h$01$^m$58$^s$.8, DEC$_{J2000}$=-63\deg
58$^\prime$10$^\prime$$^\prime$ (the conservative error estimation is
3$^\prime$$^\prime$).  This position is about 6$^\prime$$^\prime$ from
the best {\ROSAT\ } position of \rxp.  The uncertainty of the
localisation of \rxp~ with {\ROSAT\ } is 5$^{\prime\prime}$ (ROSPSPC
catalog\footnote{ftp://ftp.xray.mpe.mpg.de/rosat/catalogues/2rxp/pub}),
therefore we  conclude that most likely \xmm\ source 
and the \ROSAT\ one are the same.

\subsection{Spectral Analysis}

\begin{figure}
\begin{center}
\includegraphics[width=8cm,angle=0]{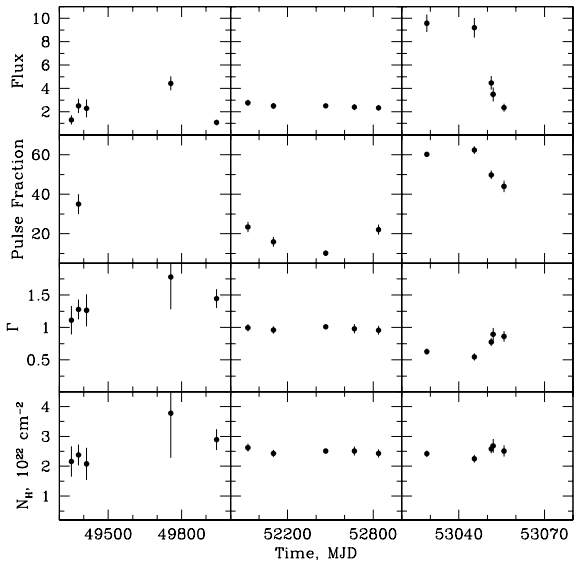}
\end{center}
\caption{Time evolution of the spectral parameters of \rxp\ and  $2-10$ keV pulse
 fraction (in \%). Flux is given in units of $10^{-11}$ erg/s/cm$^2$. }
 \label{spcpar}
\end{figure}

The 1993--2004 time history of the $2-10$ keV flux from \rxp\ as observed by 
\asca\  
and \xmm\ is shown in the upper panel of
Fig. \ref{spcpar}. While during the \asca\ and the first half of the \xmm\  observations
(X1 -- X5) the flux of the source was practically
constant at a value $\sim 2.5\times10^{-11}$ ergs cm$^{2}$ s$^{-1}$,
an outburst can be seen between the end of January and the beginning of February 2004 (obs. X5 --
X10). During this period the source flux increased by a
factor of more than 5. Due to the  strategy chosen for the
\psrb\ monitoring campaign, the whole outburst was not entirely
covered.  During the flare \rxp\ was observed only twice (24 Jan and
10 Feb), with approximately  the same flux level. During the
following 10 days its flux dropped   to the 2001-2003 level with a
characteristic decay time of $\sim7.5$ days (Fig.~\ref{spcpar}).

As  can be seen from Table \ref{intdata}, this outburst was also
detected by \intgr\ in the $20-60$ keV energy range.  While in 2003
 out of a  $\sim260$ ks observation, the source was only marginally 
detected at $\sim3\sigma$ level, it was clearly seen
during a  6.1 ks observation on January 26, 2004 (I3), and during  
a 4.2 ks observation on February 7, 2004 (I4). At those times it was 
ten times brighter than its averaged level over 2003. On February 19, 
2004 (I5) and during a ToO observation of \psrb\ performed in March 
2004 \citep{shaw}, the source was again only marginally or not detected.

\begin{figure}
\begin{center}
\includegraphics[width=8cm,bb=0 100 666 666]{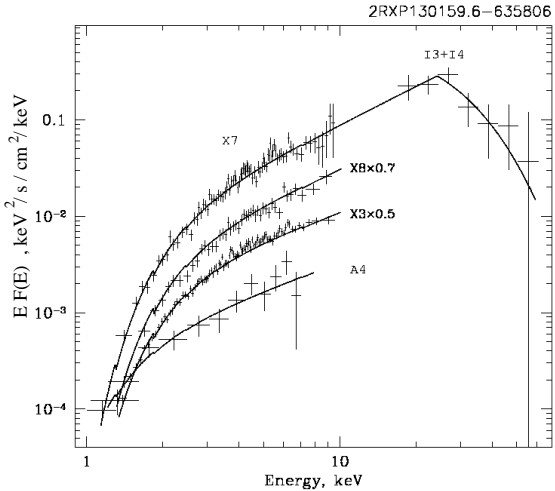}
\end{center}
\caption{ Spectral evolution of \rxp, as observed with \xmm\, \asca\
and \intgr.  To better show the spectral evolution for the \xmm\
observations the spectra from obs. X3 and obs. X8 were multiplied by 0.5 and
0.7 respectively. The combined \xmm\ and \intgr\ spectrum is
fitted with an absorbed  power law model with a high energy cutoff.}
\label{spectry}
\end{figure}

The spectral analysis was done with the XSPEC software
package. The spectrum of \rxp~ during the lowest
state as observed with \asca\ in 1994 (obs. A4), a typical \xmm\ spectrum of
the source in 2002 -- 2003 (obs. X3), \xmm\ and \intgr\ spectra during the
outburst (obs. X7, obs. I3+I4) and just after (obs. X8), are shown on  
Fig.~\ref{spectry}.

The \xmm\ and \asca\ data show that the spectrum of the source in the soft
$2-10$ keV energy range is well described by a simple power law
modified by absorption. In Table \ref{fitpar} we present results of
three-parameter fits.  The uncertainties are given at the $1\sigma$
statistical level and do not include systematic uncertainties. A
graphical representation of the evolution of the spectral parameters
is shown in Fig.~\ref{spcpar}.


For all observations, the value of the photo-absorption is
practically constant with an average value of
$N_{\mathrm{H}}=(2.48\pm0.07) \times 10^{22}$cm$^{-2}$. This value is
about five times higher than the value we found for \psrb\ ($0.48\pm0.03
\times 10^{22}$cm$^{-2}$), which is located only 10 arcminutes away
(Chernyakova et al., 2005b).  Measurements of the interstellar hydrogen 
in the Galaxy by Dickey \& Lockman (1990) give $N_{\mathrm{H}}$ values
in the range $(1.7-1.9)\times 10^{22}$cm$^{-2}$, which is smaller 
than the one we deduced from X-ray spectral fits. 
This indicates that part of the absorption might be intrinsic to the source.

While the \asca\ and \xmm\ data are well fitted with a simple power law
modified by photo-absorption (see Table \ref{fitpar}), \intgr\
data show a presence of a high-energy cut-off at about $\sim25$ keV,
which is typical for accreting X-ray pulsars \citep{white83}. We fitted the
joint spectrum obtained with \xmm\ (X7) and \intgr\ (I3+I4) with an absorbed 
cut-off power law. The best fit parameters obtained are: $N_{\mathrm{H}}=(2.55\pm0.13)
\times 10^{22}$cm$^{-2}$, $\Gamma=0.69\pm0.05$, $E_{cut}=24.3\pm3.4$
keV, $E_f=8.5\pm3.3$ keV. The normalization of the INTEGRAL/IBIS
spectrum was taken as arbitrary.

\begin{table}
\caption{Models Parameters for \asca\ and \xmm\ observations of \rxp.}\label{fitpar}
\begin{tabular}{c|c|c|c|c}
\hline
Data& N$_{\mathrm{H}}$&              Photon  & Flux$_{2-10}^*$         &$\chi^2$ (dof) \\
 Set& 10$^{22}$ cm$^{-2}$& Index         & &             \\
\hline
A1&  2.16 $\pm$ 0.51&   1.11 $\pm$ 0.22&   1.30 $\pm$ 0.40 &  0.86 (187) \\
A2&  2.38 $\pm$ 0.35&   1.28 $\pm$ 0.15&   2.51 $\pm$ 0.62 &  0.97 (187) \\
A3&  2.08 $\pm$ 0.54&   1.26 $\pm$ 0.25&   2.29 $\pm$ 0.76 &  0.66 (187) \\
A4&  3.78 $^{+1.59}_{-1.27}$&   1.78 $\pm$ 0.50& 4.43 $^{+0.67}_{-0.56}$ &  0.95 (198) \\
A5&  2.89 $\pm$ 0.35&   1.45 $\pm$ 0.15&   1.08 $\pm$ 0.24 &  1.00 (760) \\
X1&  2.62 $\pm$ 0.13&   1.00 $\pm$ 0.06&   2.77 $\pm$ 0.27 &  1.00 (255) \\
X2&  2.43 $\pm$ 0.12&   0.96 $\pm$ 0.06&   2.50 $\pm$ 0.25 &  1.03 (245) \\
X3&  2.51 $\pm$ 0.06&   1.01 $\pm$ 0.03&   2.51 $\pm$ 0.12 &  1.15 (722) \\
X4&  2.51 $\pm$ 0.15&   0.98 $\pm$ 0.07&   2.40 $\pm$ 0.26 &  0.94 (230) \\
X5&  2.43 $\pm$ 0.14&   0.96 $\pm$ 0.07&   2.34 $\pm$ 0.24 &  1.15 (241) \\
X6&  2.42 $\pm$ 0.11&   0.63 $\pm$ 0.05&   9.58 $\pm$ 0.76 &  1.19 (435) \\
X7&  2.25 $\pm$ 0.13&   0.55 $\pm$ 0.06&   9.20 $\pm$ 0.84 &  1.12 (369) \\
X8&  2.58 $\pm$ 0.15&   0.77 $\pm$ 0.06&   4.47 $\pm$ 0.58 &  1.06 (316) \\
X9&  2.68 $\pm$ 0.23&   0.90 $\pm$ 0.10&   3.49 $\pm$ 0.60 &  1.06 (173) \\
X10& 2.51 $\pm$ 0.19&   0.86 $\pm$ 0.08&   2.35 $\pm$ 0.34 &  1.00 (169) \\
\hline
\end{tabular}
$^*$ in $10^{-11}$erg cm$^{-2}$ s$^{-1}$ units.
\end{table}

\subsection{Timing Analysis}

\begin{figure}
\begin{center}
\includegraphics[width=8cm,angle=0]{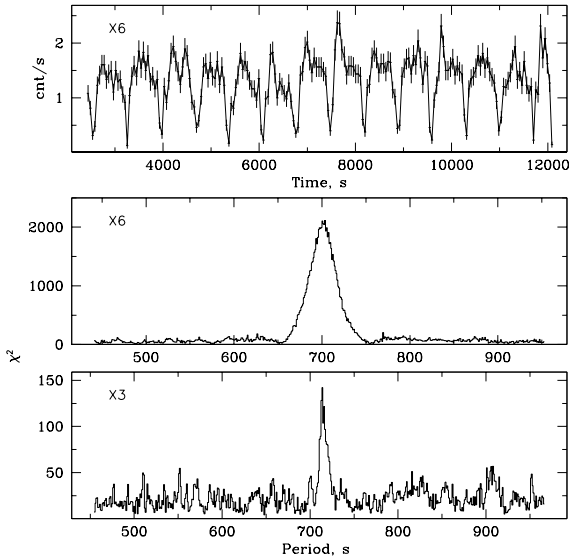}
\end{center}
\caption{MOS1 $2-10$ keV light curve of \rxp\ during the flare (obs. X6) (top panel),
and $\chi^2$ distribution versus trial period for the brightest (X6)
and the longest (X3) observations (middle and bottom panels respectively).}
\label{lcur}
\end{figure}

Analyzing  the light curve of \rxp\ obtained with \xmm\ in the bright
state we find that it demonstrates near coherent strong
variations with a characteristic time about 700 s. Fig.~\ref{lcur}
(upper panel) shows the example of a 48 s -- binned 2--10 keV MOS1 
background subtracted light curve of \rxp\  during the flare
(obs. X6).  The periodograms ($\chi^2$ distribution  versus
trial period for observations X3 and X6) are also represented in 
the same figure.  Periodic variations of the source flux
are obvious. The following analysis showed that such variations are
 also observed in the light curve of the source in low state.

\begin{figure}
\begin{center}
\includegraphics[width=7.5cm,bb=40 180 565 720,angle=0]{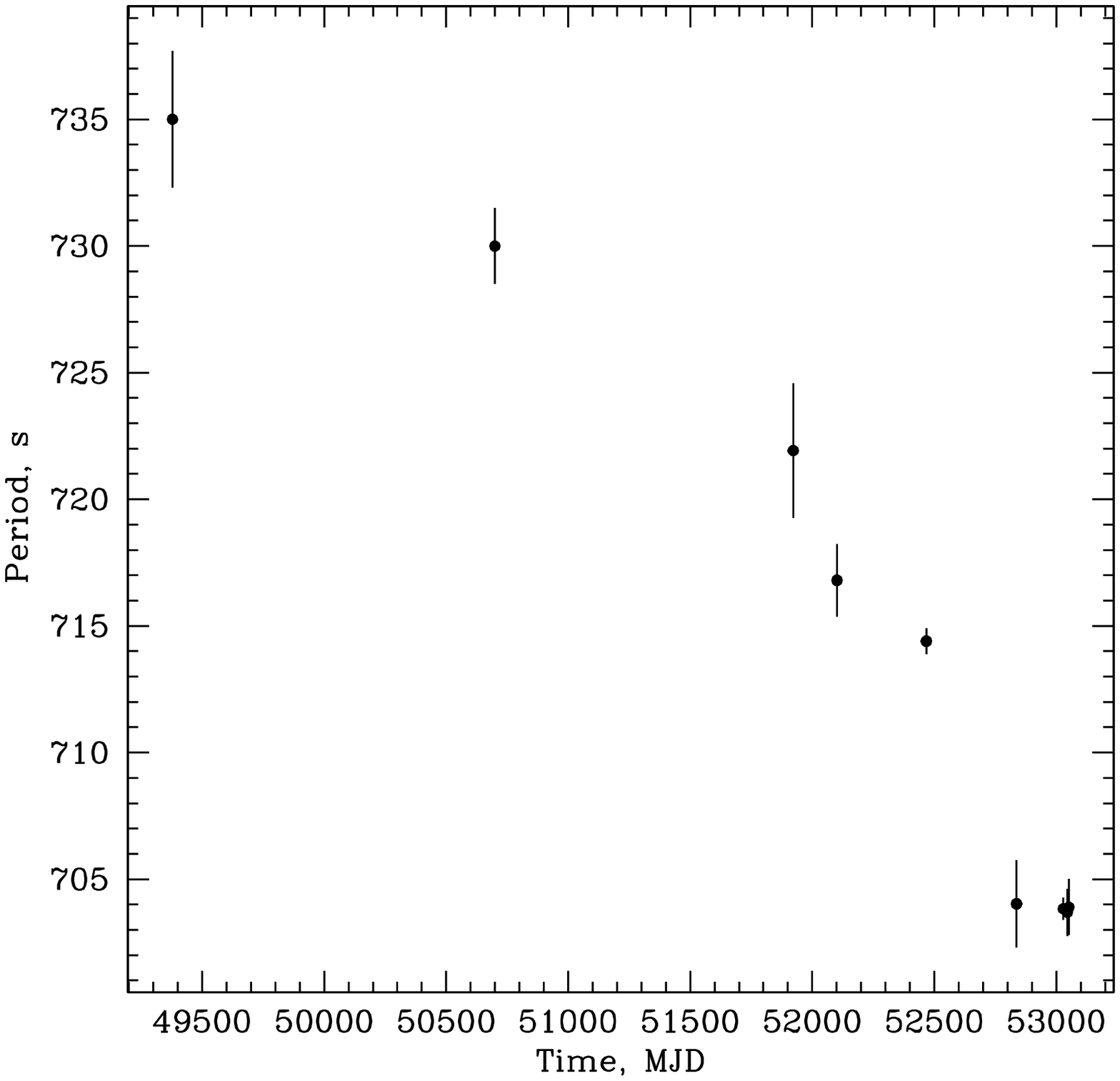}
\end{center}
\caption{Time evolution of \rxp\ pulse period. }
\label{histper}
\end{figure}
\begin{figure}
\begin{center}
\includegraphics[width=8cm,bb=80 180 460 720,angle=0]{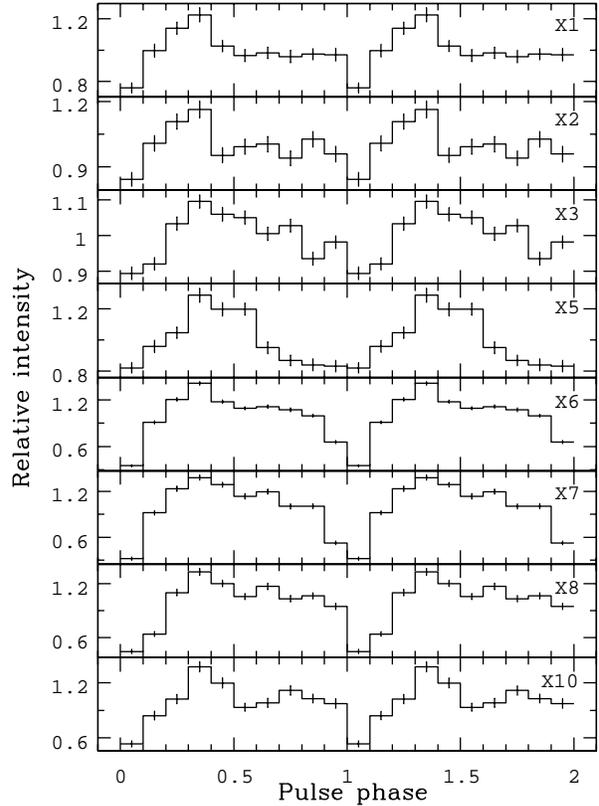}
\end{center}
\caption{\rxp\ pulse profiles variations in the $2-10$ keV energy range. Pulse
profiles have been aligned using the minimum phase bin.}
\label{pprof}
\end{figure}

Subsequent analysis of \asca\ and \sax\ light curves of the source flux also 
showed pulsations, although  not in all the datasets. This is  due to 
the much 
smaller statistics of these data. With  \intgr\ data we can set only  upper
limit on the pulse fraction. For the brightest I3 and I4 observations the
  the 3 $\sigma$ upper limit is 70\%, which is consistent 
with almost simultaneous \xmm\ observations X6 and X7.

The values of the pulse period obtained between 
1994 and 2004 are given in Table \ref{pulse} and Fig. \ref{histper}.
 An average spin up rate changes from 
$\dot P\simeq
-6 \times 10^{-8}$s s$^{-1}$ in 1994 -- 2001, to approximately
$\dot P\simeq
-2 \times 10^{-7}$s s$^{-1}$ in 2001 -- 2004, that corresponds to $\dot
\nu\simeq 10^{-13}$Hz s$^{-1}$ and $\dot
\nu\simeq 4 \times 10^{-13}$Hz s$^{-1}$ respectively.

\begin{table}
\caption{\rxp\ pulse period, as observed  with \asca, \sax, and \xmm.
Observations X4 and X9 have been excluded, as they were
strongly affected by soft proton flares. \label{pulse}}
\begin{center}
\begin{tabular}{c|c|c|c}
\hline
Data&Date,&Pulse    &   Pulse     \\
 Set&MJD&Period, s&Fraction, \%  \\
\hline
A2 &49378.79 &735   $\pm$   2.7&35     $\pm$    6 \\
S1 &50699.5  &730   $\pm$   1.5&26     $\pm$    5 \\
X1 &51921.73 & 721.9$\pm$  2.7&23.4$\pm$    2.6\\ 
X2 &52101.31 & 716.8$\pm$  1.4&15.9$\pm$    2.4\\
X3 &52467.24 & 714.4$\pm$  0.5&10.1$\pm$    1.3\\
X5 &52837.53 & 704.0$\pm$  1.7&22.1$\pm$    2.5\\
X6 &53028.79 & 703.8$\pm$  0.4&60.2$\pm$    1.2\\
X7 &53045.43 & 703.7$\pm$  0.9&62.3$\pm$    1.9\\
X8 &53051.39 & 703.9$\pm$  1.1&49.8$\pm$    2.2\\
X10&53055.82 & 704.2$\pm$  1.1&44.0$\pm$    2.9\\
\hline
\end{tabular}
\end{center}
\end{table}
The 2--10 keV pulse profiles of \rxp\ obtained in 
each data set by folding of the \xmm\ light curves at the  
best fitted period are shown in Fig.~\ref{pprof}. In general 
the source pulse profile consists of one
broad peak, but in several observations (the low intensity ones) some
additional features (as  a second peak) are visible. We have calculated
the $2-10$ keV pulse fraction 
$P=(I_{max}-I_{min})/(I_{max}+I_{min})$ (where $I_{max}$ and $I_{min}$
are intensities at the maximum and minimum of the pulse profile) in
all the \xmm\ observations. These values are plotted on 
Fig.~\ref{spcpar} (second panel from the top). It is interesting to note that 
the pulse fraction is not constant and varies with time from  
$\sim$10 -- 25\% to $\sim60$ \% during the outburst.

\begin{figure}
\begin{center}
\includegraphics[width=8cm,bb=20 280 550 720,angle=0]{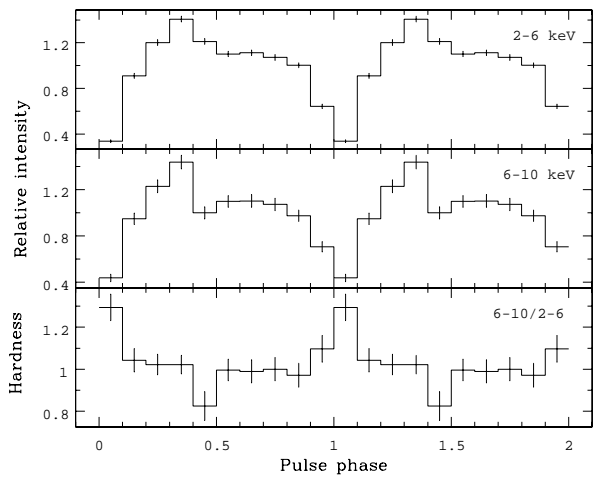}
\end{center}
\caption{$2-6$ keV and $6-10$ keV pulse profiles of \rxp\
during the brightest observation (X6) along with the hardness ratio. }
\label{pprof2610}
\end{figure}

Fig.~\ref{pprof2610} shows $2-6$ keV and $6-10$ keV pulse profiles of
\rxp\ during the brightest observation (obs. X6) along with the hardness ratio. We can 
see that the hardness remains practically constant during the pulse,
except just before phase 0.5, where it suddenly drops by
$\sim$20\%, and near the pulse minimum (around phase 1), where it
increases by $\sim$20\%. 

\begin{figure}
\begin{center}
\includegraphics[width=8cm,bb=20 50 500 350,angle=0]{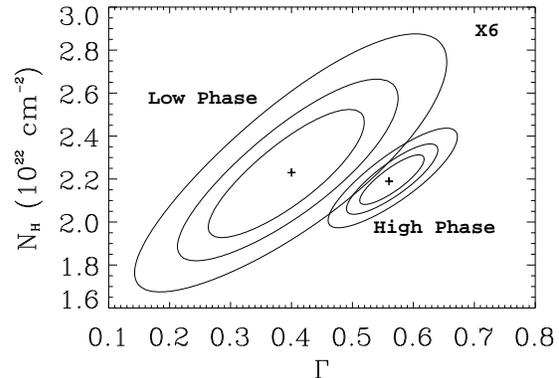}
\end{center}
\caption{ Confidence contour plots of the column density $N_H$ vs. photon
 index $\Gamma$  for a power-law fit to high and low phases of  obs. X6.
 The contours are the  68\%, 90\%, and 99\% confidence levels.}
\label{cntr}
\end{figure}

 In order to study the reasons of the variations in the shape of the 
 pulse profile, we extracted separately spectra of obs. X6  
from the low phase and from high phase. The background spectra were extracted 
from the same GTIs,  and response files were produced as explained in Section \ref{xmmobs}.
We then fitted the resultant spectra in XSPEC with a simple model of an absorbed
power law. The best fit parameters are $N_{\mathrm{H}}=2.2\pm0.3\times 10^{22}$ cm$^{-2}$, and
$\Gamma=0.38\pm0.14$ with a 2-10 keV unabsorbed flux of $6.9 \times 10^{-11}$ erg cm$^{-2}$ s$^{-1}$
for the low phase, and $N_{\mathrm{H}}=2.19\pm0.12\times 10^{22}$ cm$^{-2}$, and
$\Gamma=0.56\pm0.05$ with a 2-10 keV unabsorbed flux of $1.5 \times 10^{-11}$ erg cm$^{-2}$ s$^{-1}$
for the high phase. In both cases the reduced $\chi2$ is close to 1.
The $\Gamma$-$N_{\mathrm{H}}$ contour plots for both phases are given  on Fig. \ref{cntr}. 
It is clear that the variations between both phases are not due to variations of the absorbing 
column density.
On the contrary they seem to reflect some changes in the spectral properties of the emitting
medium, since there is some increase in the photon index. However at the 3 $\sigma$ level
both values are still compatible. 

We investigated the energetic dependence of the pulse fraction for 
the bright state of the source (obs. X6) and found that
it is more or less stable around $53$--$55$\% in the 4-10 keV energy
band and increases to $\sim63$\% in the soft 2-4 keV band.

\section{Optical counterpart and the source distance}

The accretion powered X-ray pulsars are usually found
within high-mass X-ray binaries (HMXB). The HMXB may be divided mainly 
into those with  main-sequence Be star companions, and those with
evolved OB supergiants companions.

In the case of Be/X-ray binaries the
hard X-ray emission
is caused by accretion of circumstellar material on to the neutron star.
The source of accreting material is thought to be concentrated
 towards the equatorial
plane of the rapidly rotating Be stars. Most of Be/X-rays binaries are
transients, displaying X-ray outburst and long period of quiescence, when no
X-ray flux is detected.
 A smaller group of Be/X-rays binaries
are persistent sources with rather low X-ray luminosity ($<10^{35}$erg/s),
 relatively long ($>200$s) pulse periods and very weak iron line at 6.4 keV.
(Reig \& Roche, 1999, Negueruela 2004, Haberl \& Pietsch 2004). 
  

The supergiant binaries may be further subdivided into two classes,
depending on whether the mass transfer is due to the Roche lobe overflow,
or a capture from the stellar wind. As the typical spin period for the
pulsars with the companions filling its Roche lobe is less than 20 seconds
(e.g. Corbet 1986) such a companion seems to be unlikely
for \rxp. The wind-fed  supergiant binaries has  long (of several hundreds
seconds) spin period, and are pesistent sources with short,
irregular outbursts(e.g. Corbet 1986, Bildsten et al. 1997, Negueruela 2004).
All the known systems display approximately the same X-ray luminosity 
$\sim 10^{36}$ erg/s. Variable X-ray activity of \rxp\ indicates that 
this binary system unlikely contains an OB supergiant.

In any of the cases mentioned above we should expect that the optical
companion of the X-ray source should be bright in the optical and infrared
spectral bands. In order to check this we used the results of DSS and
2MASS surveys. In both
catalogs a relatively bright star with magnitudes $B=17.2$, $R=13.9$,
$J=8.87$, $H=7.53$, $K=7.01$, is visible in the vicinity of the X-ray
source, but its position is just outside the \xmm\ error box (the
offset between the positions is $\sim4.4$$^\prime$$^\prime$). Besides
this bright star another possible counterpart candidate is found in
2MASS with coordinates (equinox 2000) RA=13$^h$01$^m$58$^s$.7,
DEC=-63\deg 58$^\prime$09$^\prime$$^\prime$ (at
$\sim 1.1$$^\prime$$^\prime$ from the best \xmm\ position) and
magnitudes $J=12.96\pm 1.33$, $H=12.05\pm0.03$,
$K_s=11.35\pm0.09$. The good agreement between both positions would
tend to suggest that this second source is the likely counterpart to
\rxp.

To estimate the de-reddened magnitude we assume that this
counterpart was only absorbed by the Galactic absorption on the line
of sight. Using the value of $N_{\mathrm{H}} =1.7
\times10^{22}$ cm$^{-2}$  we estimate the de-reddened magnitudes
$J_{der}=10.73\pm1.33$, $H_{der}=10.72\pm0.03$,
${K_s}_{der}=10.51\pm0.09$ (only statistical uncertainties are quoted).
 If the companion star is a Be main sequence star
with surface temperature around 10000 K and the radius around
6-10 $R_{\odot}$ we can expect to see its infrared brightness $J, H,
K\sim10-11$ if the binary system is at the distance $\sim 4-7$ kpc.
An additional tentative argument in favour of
such source distance is the source location in the direction to the Crux
spiral arm tangent, as HMXBs are concentrated towards galactic spiral arms
\citep{grimm02,lut05:0}.  At such a distance  unabsorbed intrinsic luminosity of  
\rxp\ is about $\sim 5\times 10^{34}$ -- $10^{35}$ erg/s, \textit{i.e.} compatible with the typical
luminosities of the persistent Be/X-rays binaries.

\section{Conclusions}

We report the identification by \xmm\ of a new X-ray pulsar with a
spin period of $\sim700$ s in the region of the Crux spiral arm.
The source was observed several times in 1993-2004 with \asca, \sax\ and
\xmm\ during the monitoring campaigns of the pulsar \psrb. The typical
flux measured from the source in the $2-10$ keV energy band is about
$(2-3)\times 10^{-11}$ erg cm$^{-2}$ s$^{-1}$, but in Jan-Feb 2004
an outburst with more than 5 times increase of the intensity was
observed from the source. During this outburst the source was also
detected in the hard X-rays with the \intgr\ observatory. Strong
pulsations of the X-ray flux with a period $\sim700$ s were detected. 
The study of a set of 
observations has shown that the pulse period changed from $\sim735$ sec
in 1994 to $\sim 704$ sec in 2004. The
average value of the spin-up rate is
$\dot \nu\simeq 2\times 10^{-13}$Hz s$^{-1}$, that is typical for
accretion powered X-ray pulsars (see e.g.  Bildsten et
al. 1997). Long pulsation period indicates that the pulsar likely
resides in a binary system with a massive companion.
The proposed infrared counterpart to the X-ray source does not contradict 
this hypothesis. From  brightness of the infrared counterpart measured, 
a tentative estimate of the distance of binary system  is 4-7 kpc, which
can indicate that the HMXB is located close to the Crux spiral arm tangent.

\section{Acknowledgements}
The authors acknowledge useful discussions with  T.J.-L. Courvoisier, P. Hakala, A. Paizis, 
I. Kreykenbohm, S.E. Shaw,  
and thank L. Sidoli for valuable comments on the details of \sax\ data reduction.
Authors are grateful to L. Foschini for the helpful advises on the \xmm\ 
data analysis. 
Authors thank  S. Molkov for the support of this work. Authors are 
grateful to the anonymous 
referee for helpful comments.
This work was partially done during AL visits to the INTEGRAL Science
Data Centre. AL thanks the ISDC staff for its hospitality and computer
resources; these visits were supported by ESA. AL also acknowledges
the support of RFFI grant 04$-$02$-$17276.

\label{lastpage}

\end{document}